\documentclass[]{aastex631}

\begin{document}

\title{Constraints on Europa's water group torus from HST/COS observations}

\author{Lorenz Roth}
\affiliation{Space and Plasma Physics, KTH Royal Institute of Technology, Stockholm, Sweden}

\author{H. Todd Smith}
\affiliation{The Johns Hopkins University Applied Physics Laboratory, Laurel, MD, USA}

 \author{Kazuo Yoshioka}
 \affiliation{University of Tokyo, Chiba, Japan.} 
 
\author{Tracy M. Becker}
\affiliation{Southwest Research Institute, San Antonio, TX, USA}

\author{Aljona Blöcker}
\affiliation{Department of Earth and Environmental Sciences, Ludwig-Maximilians-University, Munich, Germany}

\author{Nathaniel J. Cunningham}
\affiliation{Department of Physics and Astronomy, Nebraska Wesleyan University, Lincoln, NE, USA}

\author{Nickolay Ivchenko}
\affiliation{Space and Plasma Physics, KTH Royal Institute of Technology, Stockholm, Sweden}

\author{Kurt D. Retherford}
\affiliation{Southwest Research Institute, San Antonio, TX, USA}

\author{Joachim Saur}
\affiliation{Institut für Geophysik und Meteorologie, Universität zu Köln, Köln, Germany}

\author{Michael Velez}
\affiliation{University of Texas at San Antonio, San Antonio, TX, USA}
\affiliation{Southwest Research Institute, San Antonio, TX, USA}

\author{Fuminori Tsuchiya}
\affiliation{Tohoku University, Sendai, Japan}



\begin{abstract}
In-situ plasma measurements as well as remote mapping of energetic neutral atoms around Jupiter provide indirect evidence that an enhancement of neutral gas is present near the orbit of the moon Europa. Simulations suggest that such a neutral gas torus can be sustained by escape from Europa's atmosphere and consists primarily of molecular hydrogen, but the neutral gas torus has not yet been measured directly through emissions or in-situ. Here we present observations by the Cosmic Origins Spectrograph of the Hubble Space Telescope (HST/COS) from 2020 and 2021, which scanned the equatorial plane between 8 and 10 planetary radii west of Jupiter. No neutral gas emissions are detected. We derive upper limits on the emissions and compare these to modelled emissions from electron impact and resonant scattering using a Europa torus Monte Carlo model for the neutral gases. The comparison supports the previous findings that the torus is dilute and primarily consists of molecular hydrogen. A detection of sulfur ion emissions radially inward of the Europa orbit is consistent with emissions from the extended Io torus and with sulfur ion fractional abundances as previously detected.  
\end{abstract}



\section{Introduction}

The volcanic moon Io was long thought to be the only essential source of material for Jupiter’s magnetosphere \citep{thomas04}. A neutral or plasma torus near the moon Europa, orbiting Jupiter at 9.4 R$_\mathrm{J}$ (Jupiter radius, 1~R$_\mathrm{J}$ = 71~200~km), was the subject of speculation, based on plasma characteristics near Europa measured by the Voyager spacecraft \citep{Intriligator1982,schreier93}, but could not be conclusively derived. Two decades after Voyager, two spacecraft measurements independently suggested the presence of an additional neutral gas torus in Jupiter’s magnetosphere roughly centered at Europa’s orbit: The Ion And Neutral Camera (INCA) onboard the Cassini spacecraft measured an enhanced flux of energetic neutral atoms (ENAs) that originated from a distance of 9--10 R$_\mathrm{J}$ from the planet \citep{mauk03}. At similar radial distances, a depletion of energetic protons was detected by the Energetic Particles Detector (EPD) on board the Galileo spacecraft persistently over seven years \citep{lagg03}. Both observations can be explained by charge exchange reactions between energetic plasma ions and cold neutrals from a torus near Europa’s orbital distance, which would lead to a depletion of protons and the generation of ENAs. \textcolor{black}{While a proton depletion can be caused by different effects, the particular pitch angle dependency of the detected depletion suggested a neutral gas torus as cause.} The estimated neutral densities in the equatorial plane are around 20--50~cm$^{-3}$ for total torus gas contents of several times 10$^{33}$ particles \citep{lagg03,Mauk2004}. However, ultraviolet observations by Cassini did not detect emissions from oxygen near Europa's orbit hinting that \textcolor{black}{not oxygen but} hydrogen, which is more difficult to detect, is the main species of the gas torus \citep{hansen05}.    

\textcolor{black}{The nature of a Europa neutral torus source was also studied through simulations of the atmosphere, plasma interaction, escape and torus.} Early on, \cite{saur98} estimated a neutral O$_2$ loss rate from Europa of 50 kg/s through ion-neutral collision including charge-exchange reactions, which could possibly sustain an oxygen torus. The neutral particle simulations by \citep{shematovich05,smyth06} concluded that loss from the global atmosphere is likely a sufficient source to maintain a neutral torus in Europa's orbit and that the main constituent of the torus should be molecular hydrogen. \cite{smyth06} found atmosphere loss rates to the torus of 9.3 kg/s for H$_2$ and 8.0 kg/s for O. A combined atmosphere-chemistry and magnetohydrodynamic interaction model by \cite{dols16} suggested that charge-exchange cascades are the main process leading to escape of O$_2$ from the atmosphere to the torus, resulting in a dilute but widely extended distribution of the oxygen. The possible presence of outgassing at plume locations as reported \citep[e.g.,][]{roth14-science,jia18} might be an additional source for a torus, although estimated plume gas ejection velocities are well below the escape velocity \citep{roth14-apocenter} in contrast to Enceladus \citep{tian07,Hansen2020}. 

A three-dimensional simulation of the generation and fate of a neutral torus together with a detailed analysis of the ENA profiles \citep{smith19} provided global distributions in the Jovian magnetosphere, which suggest that the neutral torus is highly asymmetric but with average densities as indicated by the observations \citep{Mauk2004,lagg03}. The study also finds H$_2$ to be the main constituent, followed by H, O, and O$_2$ in order of abundance \citep[see, e.g., figure 7 in][]{smith19}.

Another indirect detection was recently provided by measurements of pick-up H$_2^+$ ions with the  Jovian Auroral Distributions Experiment (JADE) on board the Juno spacecraft \citep{Szalay2022}. Most of the neutral H$_2$ from a Europa torus is lost by ionization forming the detected H$_2^+$ ions. The ion production rate thus directly constrains the H$_2$ torus source rate from Europa, which is estimated to be 3--13 times lower than the simulation results from \cite{smyth06}. Although there are several independent indirect detections of a neutral population in Europa's torus, the neutral gas has not been detected through emission or absorption signatures in remote sensing observations. Signatures of extended H and O exospheres were detected up to $\sim$1~R$_\mathrm{E}$ (Europa radius, 1~R$_\mathrm{E}$ = 1680~km) for O, and $\sim$4~R$_\mathrm{E}$ for H away from Europa's limb \citep{hansen05,roth16-eur,roth17-europa,Roth2021-Eur} but not further beyond from the moon's gravity well. 

We have obtained observations with the Hubble Space Telescope's Cosmic Origins Spectrograph (HST/COS), aiming to detect faint emissions at the far-ultraviolet lines from hydrogen and oxygen at HI~1216~Å, OI~1304~Å, and OI~1356~Å. Despite the high sensitivity of COS for faint sources, the observations did not enable a detection but provided upper limits as well as a measurement of sulfur ion emissions from the extended Io plasma torus. We first describe the observations and processing of the data and obtained brightnesses. Thereafter, we explain the modeling used to estimate expected emission brightnesses for neutrals and sulfur ions at the observed wavelengths. Finally, we compare our upper limits on H and O emissions and the measured sulfur ion brightnesses to the modeled brightnesses and discuss the results. 

\section{HST/COS observations}

\begin{figure}[htb]
\begin{center}
\includegraphics[width=0.9\textwidth]{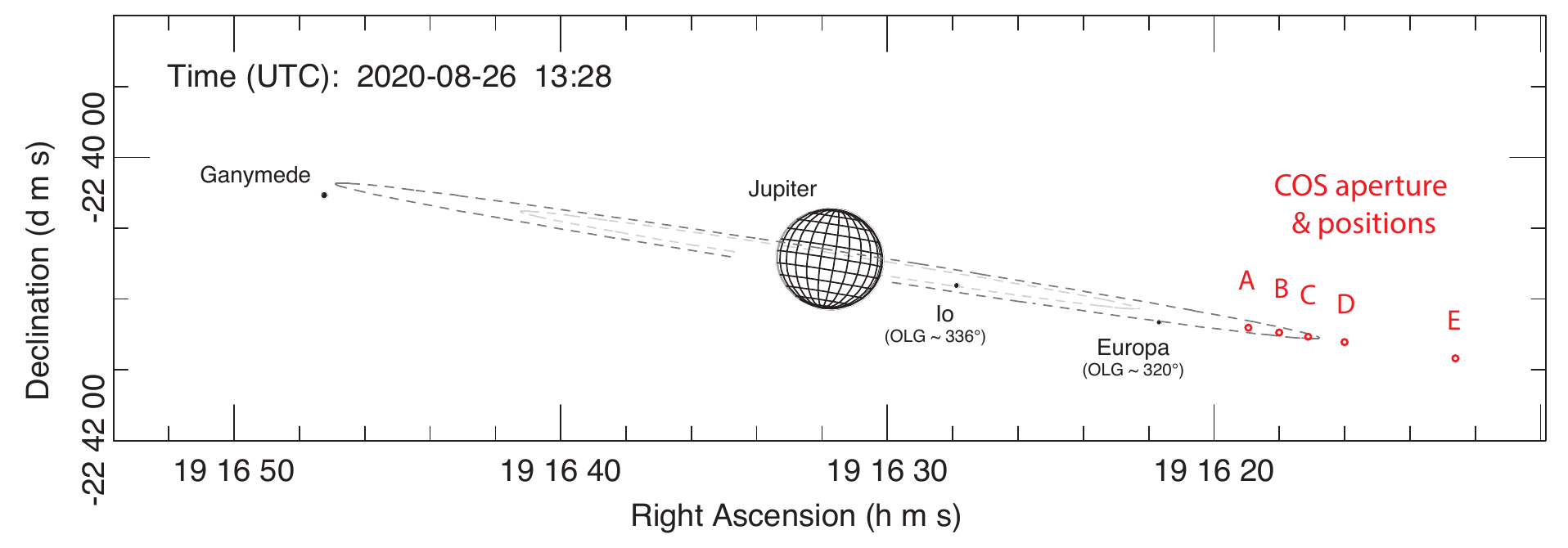}
\caption{Observing geometry showing the size and scanning positions of the round 2.5"-wide COS aperture as well as the Galilean moons at their orbital position at the start of the first exposure from 2020 (le2o01010, Table \ref{tab:obs}). Callisto is beyond the field. The grey dashed lines indicate the moon orbits for Io and Europa, which are both past Western elongation approaching eclipse (i.e., moving leftward) \textcolor{black}{(Adapted from \url{https://pds-rings.seti.org}.)}
\label{fig:obsgeom}}
\end{center}
\end{figure}

HST/COS was used with the G130M grating centered at 1291~Å and the Primary Science Aperture (PSA) during two visits in 2020 and 2021. The spectral setup provides wavelength coverage of about 1290-1430~Å on detector segment~A and 1135-1275~Å on segment~B with a spectral dispersion of 0.01~Å per pixel \citep{Hirschauer2021}. During both visits, the COS aperture was centered on Europa's orbital plane scanning over different radial distances from Jupiter's center from 8.0~R$_\mathrm{J}$ (named position 'TORUS-A', closest to the Io torus) out to 12.5~R$_\mathrm{J}$ ('TORUS-E'). Position TORUS-C was centered on the radial distance of 9.3~R$_\mathrm{J}$ just inside the orbital distance of Europa, where highest emissions were expected \citep{smith19}. In \textcolor{black}{2020}, one exposure each was taken at two intermediate distances of 8.6~R$_\mathrm{J}$ ('TORUS-B') and 10.0~R$_\mathrm{J}$ ('TORUS-D') in addition. Figure \ref{fig:obsgeom} illustrates the complete scan from 2020, with the moons' orbital positions shown at the start. Table \ref{tab:obs} provides an overview on the exposures including the pointing position. 

\begin{table}[htb]
\caption{HST/COS observations of the Europa torus}
\label{tab:obs}
\resizebox*{1.\textwidth}{!}{
\begin{tabular}{llcccccccc}
\hline
Exp. ID     & Pointing -    & Date 		&   Start	& Total      & Used time  & Used time  & Europa   & Io         & Jup. SysIII	\\
 			& Radial distance	&           &	time	& exp.time   & for O \& S     & for H~Ly-$\alpha$ & orb.lon$^a$  & orb.lon$^a$  & longitude$^{a,b}$	\\ 
			&		to Jupiter	&            &	(UTC)    & 	 [sec]   & [sec]      & [sec]  & [$^\circ$] & [$^\circ$] & [$^\circ$]	\\		
\hline
\textbf{2020} \\
\hline
le2o01010   &   TORUS-E -- 12.5~R$_\mathrm{J}$ &   2020-08-26  & 13:28 &   2611 & 1611 &   750 & 322 & 340 & 279 \\
le2oa1010   &   TORUS-C -- 9.3~R$_\mathrm{J}$&   2020-08-26  & 15:05 &   2699 & 1799 &   750 & 329 & 354  & 337  \\
le2o02010   &   TORUS-B -- 8.6~R$_\mathrm{J}$&   2020-08-26  & 18:20 &   2611 & 1911 &   750 & 342 &  21 & 94 \\
le2oa2010   &   TORUS-D -- 10.0~R$_\mathrm{J}$&   2020-08-26  & 19:50 &   2699 & 1699 &   750 & 349 &  34   & 150 \\
le2oa3010   &   TORUS-A -- 8.0~R$_\mathrm{J}$&   2020-08-27  & 00:36 &   2699 & 1699 &   750 & 9   &  74  & 323 \\
le2oa4010   &   TORUS-E -- 12.5~R$_\mathrm{J}$&   2020-08-27  & 05:22 &   2699 & 1699 &   750 & 29  & 115 & 136 \\
\hline
\textbf{2021} \\
\hline
le2o53010   &   TORUS-C -- 9.3~R$_\mathrm{J}$&  2021-09-07   & 05:02 &   2153 & 1753 & 1053 & 307 & 266 & 168 \\
le2o55010   &   TORUS-A -- 8.0~R$_\mathrm{J}$&  2021-09-07   & 06:38 &   2153 & 1753 & 1053 & 314 & 279 & 225 \\
le2o54010   &   TORUS-C -- 9.3~R$_\mathrm{J}$&  2021-09-07   & 08:13 &   2153 & 1753 & 1053 & 320 & 293 & 283 \\
le2ob4010   &   TORUS-E -- 12.5~R$_\mathrm{J}$&  2021-09-07   & 09:48 &   2154 & 1753 & 1053 & 327 & 306 & 341 \\
\hline
\end{tabular}
}
\tablenotetext{a}{The longitudes are stated for the midpoint of the exposures used for O \& S analysis.}
\tablenotetext{b}{Sub-observer Jovian System-III longitude (central meridian longitude of Jupiter). }
\end{table}

The observations were timed such that Europa was past Western elongation (i.e., past the largest angular separation to the right of Jupiter) at the start of each visit (see also Europa orbital longitude in Table \ref{tab:obs}). The torus brightness is expected to be highest at this geometry \citep{smith19}, but Europa itself is not within or near the aperture for any of the exposures. Due to the orbital resonance, it was inevitable that Io is also west of Jupiter in both years, which we otherwise would have avoided in order to exclude contributions from fresh Io-sourced neutrals as much as possible. 

The HST/COS spectra obtained over the full exposure time are dominated by emissions from the Earth's geocorona at HI~1216~Å (on segment B) and OI~1304~Å (on segment A), as HST is on the dayside of Earth at the beginning of the exposures in both years. A faint signal is even detected from Earth's nitrogen dayglow at the NI1200~Å line \citep{Eastes1985} in most exposures.  

\begin{figure}[htb]
\begin{center}
\includegraphics[width=0.45\textwidth]{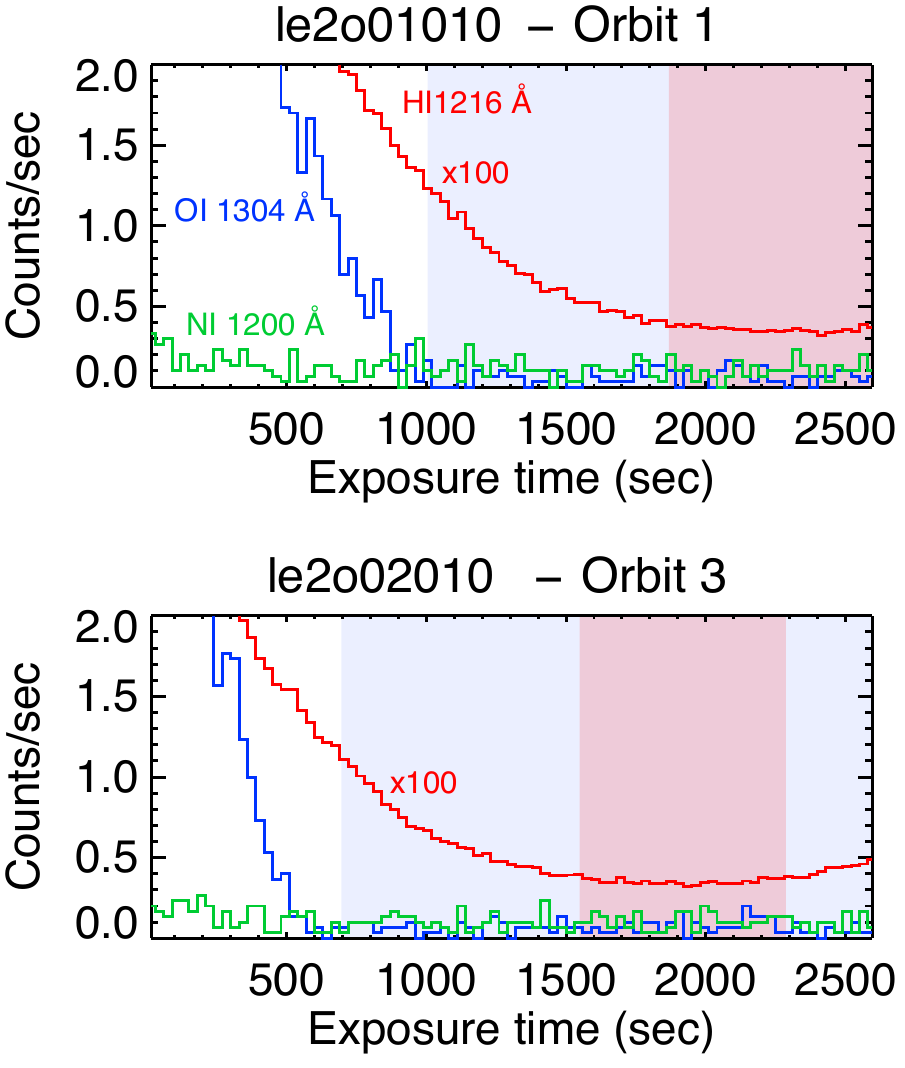}
\caption{Count rates over the exposures in 20-sec bins show the variation of geocorona emissions at OI~1304~Å, HI~1216~Å (Lyman-$\alpha$) and NI~1200~Å. The red shaded area shows the intervals used for the HI Lyman-$\alpha$ analysis, blue shaded areas are the intervals used for the neutral oxygen and sulfur ion analysis.     
\label{fig:timetag}}
\end{center}
\end{figure}

We carefully analyzed the time-variability of the geocorona emissions over the exposure in the time-tag data. \textcolor{black}{The brightness of the geocorona signal changes significantly within an exposure when HST moves from the Earth's dayside to the nightside (or vice versa).} The detector count rate near OI~1304~Å drops to a constant level after 700--1000~s in the 2020 exposures and after 400~s in the 2021 exposures. For analyzing oxygen and sulfur emissions, we used only the exposure time with the low constant OI~1304~Å count rate level, indicated by the blue shaded area for two example exposures from 2020 in Figure \ref{fig:timetag}. The geocoronal NI~1200~Å emissions are not detectable anymore in the trimmed exposures. 

The HI~1216~Å (H Lyman-$\alpha$), geocorona brightness is present throughout the exposures, slowly decreasing to a minimum near the end of the exposures, and in 2020 slightly increasing again at the end. The variation over the exposures within the visits is nearly identical in 2021 and we used the last 1053~s of each exposure (cutting 1100~s from the beginning) for the analysis of HI~1216~Å (Table \ref{tab:obs}). In 2020, the count rate variation is very similar in four of the six exposures but obviously different in the other two. The periods and profiles of minimum HI~1216~Å count rate appear to be similar, indicating that the difference originates from differing positions of HST in the Earth-Sun geometry in these two exposures. Indeed when shifting the HI~1216~Å count rate time-series curves in time relative to another, they are identical within the statistical uncertainties for all exposures. For analyzing the HI~1216~Å from the torus, we have therefore chosen an interval of 750~s around the minimum count rate, which occurs at the end during four exposures and earlier in the two other exposures (see red shaded area for the two examples in Figure \ref{fig:timetag}). 

For each exposure, the detector counts are integrated along the spatial y-axis over a height of 2.8" (28 pixels). We have tested different heights for extraction between the nominal aperture height (25 pixels) and up to 64 pixels and found that a height 28 pixels captures most of signal at a good total pixel noise level. \textcolor{black}{Note that the photon throughput outside the nominal aperture is small but non-zero so that signal can be detected even there \citep{Hirschauer2021}.} Counts are then converted to spectral brightness using the wavelength-dependent throughput of this instrument configuration. For the conversion to Rayleigh (R) we used the nominal aperture diameter of 2.5", assuming that any torus emissions are homogeneous over the aperture, which covers an area with a diameter of $\sim$8000~km at Jupiter's range (roughly 2.5 times Europa's diameter). This area is smaller than the expected scales of density gradients in neutral or plasma tori \citep{smith19,bagenaldols20}. 

To constrain emissions from a Europa torus as well as from the Io torus at all lines except H Lyman-$\alpha$, we use the spectra taken at positions TORUS-A, B, C, and D, which are considered \textit{on-target} data. The TORUS-E measurements are used for reference (\textit{off-target}) and we subtract the spectral brightness of the off-target TORUS-E exposures from the brightness measured in the on-target spectra to eliminate background and instrumental signal. After this correction, total brightnesses for specific emission lines or multiplets are calculated by integrating the spectral brightness (in R/Å) over the spectral range that the lines cover.  

For H Lyman-$\alpha$, the signal is dominated by the likely time-variable geocorona signal (even in the trimmed exposures).  We therefore compare the total brightnesses derived for all exposures for each visit (not subtracting off-target exposures for corrections). This allows us to better investigate the changes between the exposures over the visit, which are then compared to independent measurements of the solar Lyman-$\alpha$ intensity at Earth on the same days. 

In addition, the region around Lyman-$\alpha$ on the COS FUV-MAMA detector is most affected by gain-sag effects. \textcolor{black}{The high photon flux on the detector at Lyman-$\alpha$ leads to a degradation of certain detector pixels, which is difficult to assess quantitatively. Because position of the aperture on the detector and thus the overlap with gain-sag pixel areas varies from exposure to exposure, we can not exclude gain-sag regions because this would create  significant instrumental differences in Lyman-$\alpha$ flux between the exposures. Including instead all Lyman-$\alpha$ counts and investigating the variation between the exposures and including all pixels and counts, we do not need absolute flux calibration but will discuss the effects in the last section.}  

\section{Observation results}

No statistically significant emissions were detected in any of the spectra at the oxygen lines OI~1304~Å and OI~1356~Å. The derived total brightnesses in the corrected spectra (on-target spectrum minus off-target spectrum from TORUS-E position) were all consistent with zero within 2$\sigma$, i.e., smaller than two times the propagated statistical measurement error $\sigma$. 

\begin{figure}[htb]
\begin{center}
\includegraphics[width=0.5\textwidth]{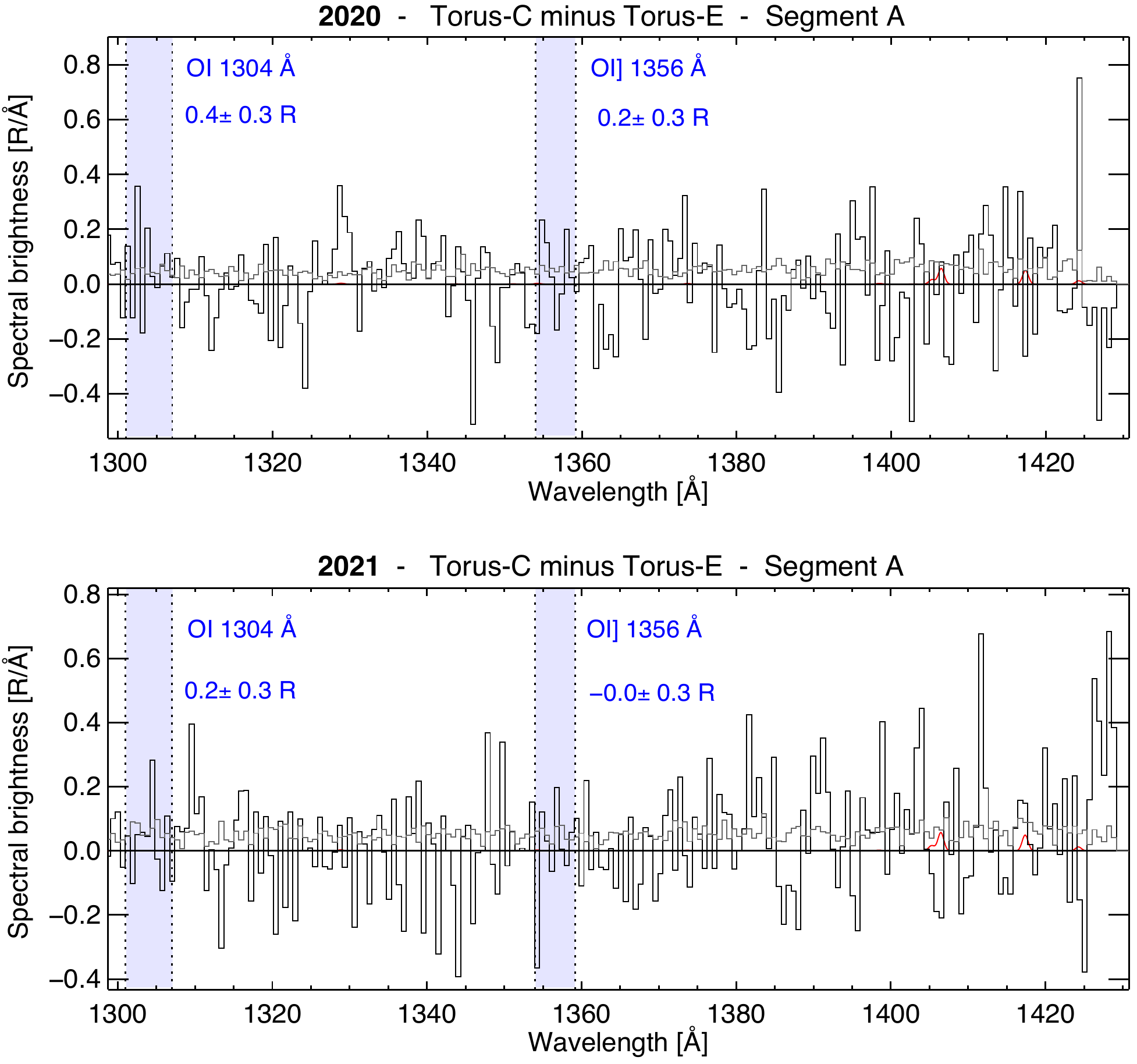}
\caption{Spectra of Torus-C position (9.3~R$_\mathrm{J}$) from 2020 (top, le2oa1010) and from 2021 (bottom, le2o54010) corrected with Torus-E spectra taken directly before and after, respectively (Table \ref{tab:obs}). The signal is shown in black, the propagated error in grey. The shaded regions show the integration boundaries for the emission brightnesses at the oxygen OI~1304~Å and OI~1356~Å lines. The modelled sulfur ion emissions at 9.3~R$_\mathrm{J}$ shown by the red peaks around 1410~Å would not be detectable. The integrated brightnesses are stated in blue with propagated uncertainties, revealing no emissions above statistical fluctuations. 
\label{fig:O_nondetect}}
\end{center}
\end{figure}

Figure \ref{fig:O_nondetect} shows two corrected TORUS-C spectra and the derived brightnesses with errors for the oxygen emissions. The propagated error in the corrected spectra is for most exposures $\sigma=0.3$~R for both oxygen multiplets, and we use 2$\sigma = 0.6$~R as upper limit constraint for the OI~1304~Å and OI~1356~Å brightnesses (Table \ref{tab:results2}).

\begin{figure}[htb]
\begin{center}
\includegraphics[width=0.5\textwidth]{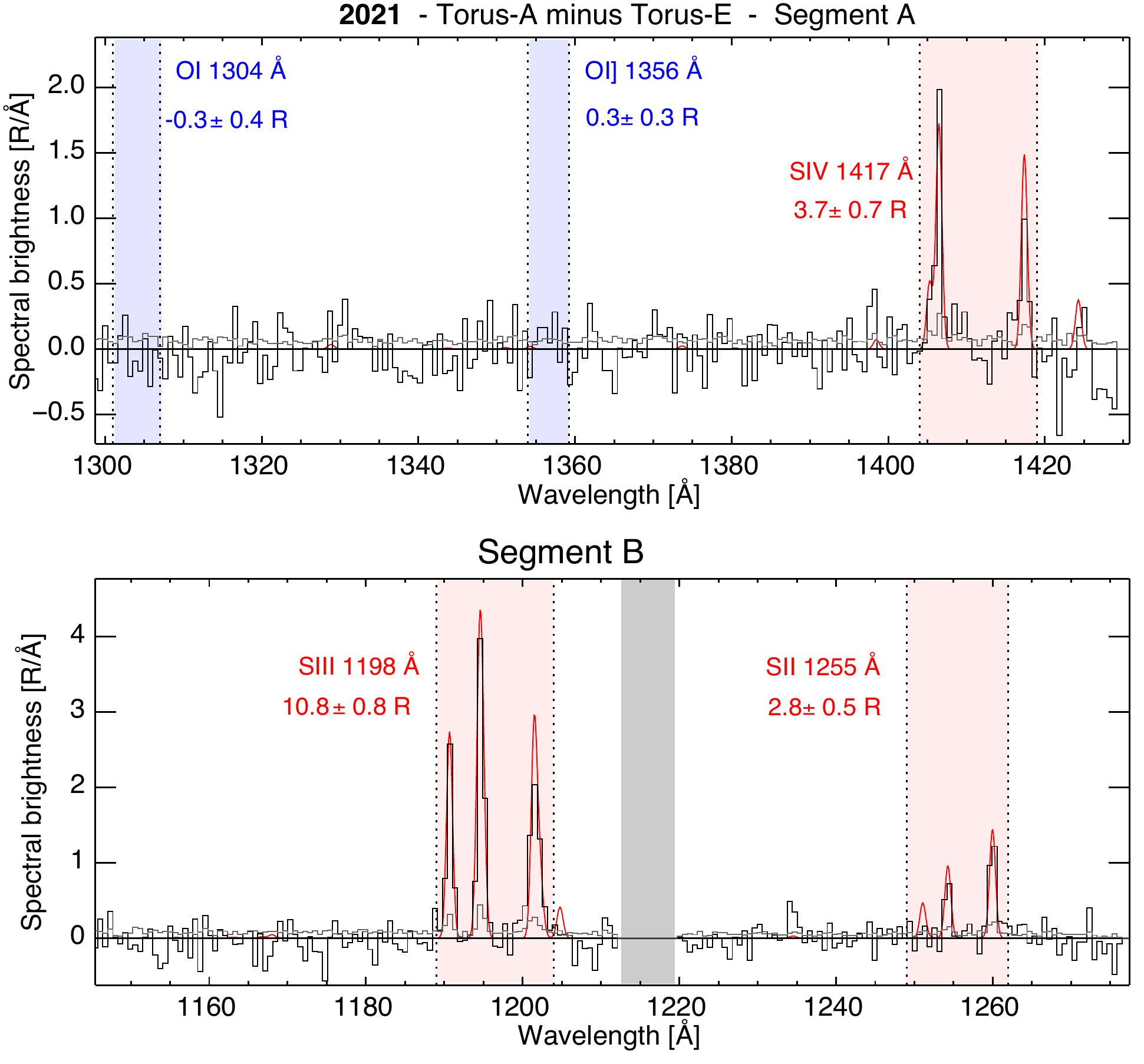}
\caption{Spectra of Torus-A position (8.0~R$_\mathrm{J}$, le2o55010) from 2021 corrected with the Torus-E spectra showing detector segment A (top) and segment B (bottom) as in Figure \ref{fig:O_nondetect}. Emission lines with significant brightnesses are detected near 1415~Å (top), 1198~Å, and 1255~Å (bottom) and identified as sulfur ion emission from Io's torus \citep[e.g.,][]{Steffl2004b}. The red shaded areas show the integration region for the ion multiplets, covering the brightest lines, and the total brightnesses with errors are stated in red. Our best-fit modelled sulfur ion emissions are shown in red. The brightness around the H Lyman-alpha line is off the scale and not shown (grey box).
\label{fig:Sion_detect}}
\end{center}
\end{figure}

In the TORUS-A exposure from the 2021 visit, statistically significant emission lines were detected near 1417~Å in segment A and near 1198~Å and 1255~Å in segment B, see red shaded and labeled areas in Figure \ref{fig:Sion_detect}. The detections are significant with signal-to-noise ratios between 6 and 14. The TORUS-A pointing is closest to the Io torus and the detected lines are identified to be sulfur ion emissions from doubly ionized sulfur (SIII) at 1198~Å, singly ionized sulfur (SII) at 1255~Å, and triply ionized sulfur (SIV) at 1417~Å \citep{morton03}. 

In the 2020 TORUS-A exposure, the brightness at these three ion multiplets is consistent with zero (no emissions) within the statistical uncertainty of about 0.3~R. Similarly, sulfur ion emissions were not detected in any of the TORUS-B, C, or D exposures in 2020 or 2021. The narrow peak at 1412~Å in Figure \ref{fig:O_nondetect} (bottom) is between the lines of the SIV multiplet, only resolution element in line width, and likely statistical. 
% We discuss the detection and non-detection further in the following section.
%

Neutral oxygen emissions from the two multiplets are not detected in the TORUS-A exposures at 8.0~R$_\mathrm{J}$ as shown in the blue areas with labels in Figure \ref{fig:Sion_detect}. The upper limits for this radially inward position are summarized and compared to the model results separately in Table \ref{tab:results2}. 

For H Lyman-$\alpha$, we find total brightnesses between 1167($\pm8)$~R (TORUS-A exposure) and 1251($\pm8)$~R (TORUS-C) in the 2020 visit, which means a variation by 84~R or 7\% of the mean. The first two exposures (le2o01010 / TORUS-E and le2oa1010 / TORUS-C) are most useful for a direct off-target to on-target comparison close in time and have almost identical brightness (difference of only 3~R ). In 2021, the brightnesses were between 1570($\pm5)$~R (second TORUS-C exposure) and 1619($\pm5)$~R (TORUS-E exposure), i.e. changes by 49~R or 3\% of the mean. 

The variations appear to be uncorrelated to the different pointing positions. For example, the signal in the on-target exposures is not systematically higher compared to the off-target exposures. Time-resolved measurements of the solar Lyman-$\alpha$ intensity by the TIMED/SEE instrument \citep{woods05} from the same day and time reveal stronger variations during the 2020 visit (19\%) than during the 2021 visit (9\%), similar to the difference in variation in the COS data (more variation in 2020). This supports the possibility that the changes between the COS exposures are due to changes in the geocoronal emissions as previously found for HST/COS observations \citep{roth18}, and not due to difference from the targeted regions. 

\begin{figure}[htb]
\begin{center}
\includegraphics[width=0.49\textwidth]{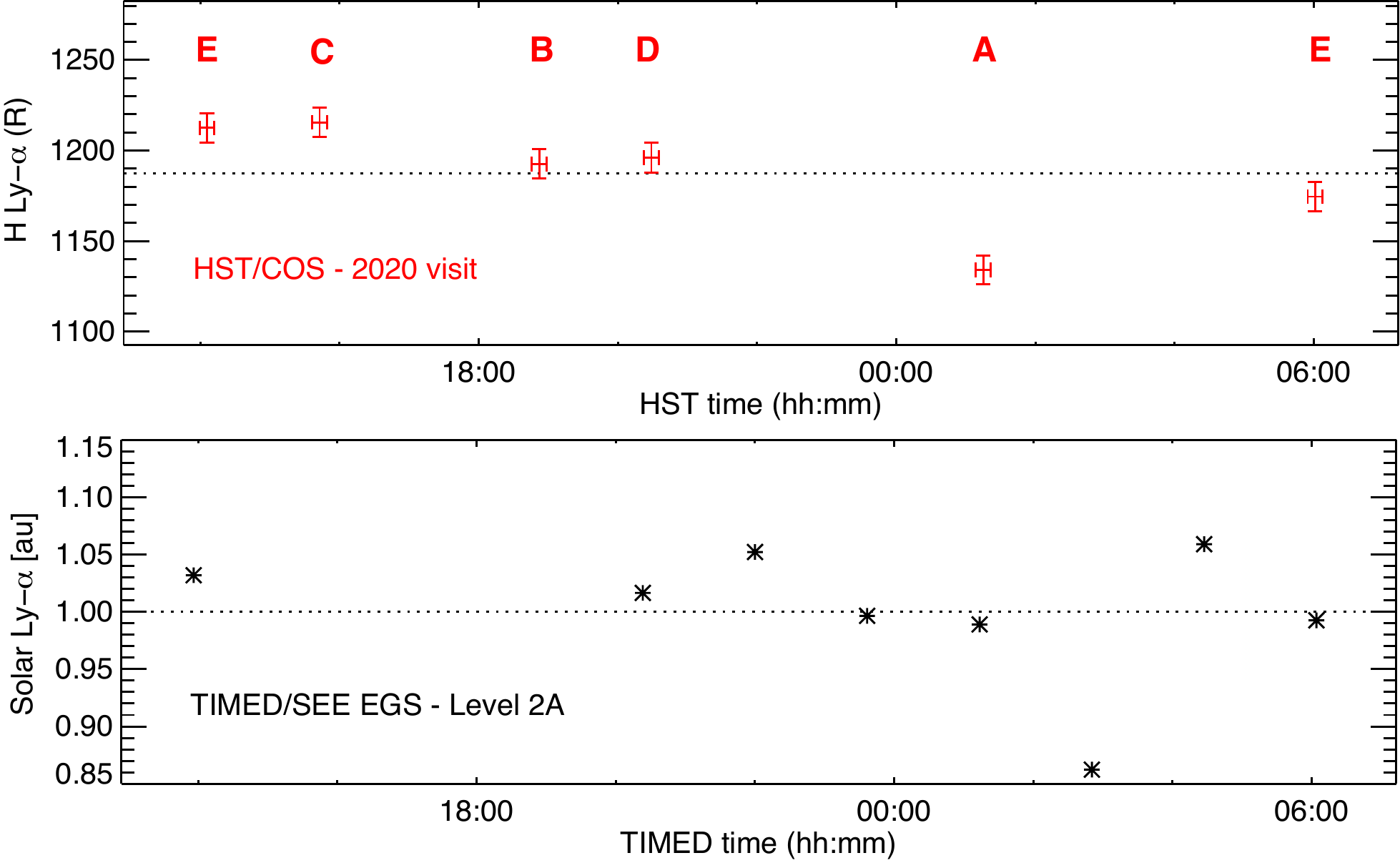}
\caption{(Top) Integrated H Lyman-$\alpha$ brightnesses from the 6 HST/COS exposures from 2020 shown at the observing time in UTC. The vertical extent of the boxes indicates the statistical uncertainties ($\sigma\sim8$~R), the width of the box illustrates the length of the exposures (750~sec each). (Bottom) Normalized Lyman-$\alpha$ flux measured by TIMED/SEE. The vertical scales cover similar ranges of $\pm$8\% around the mean brightness, which illustrates that the variance in HST/COS data is similar to the variance in the solar flux (as measured by TIMED/SEE), thus not requiring nor indicating emissions from the torus or other sources.    
\label{fig:lya}}
\end{center}
\end{figure}

Figure \ref{fig:lya} shows the Lyman-$\alpha$ brightness from the COS exposures (top panel) and the values from TIMED/SEE (lower panel, with some data gaps). There is no clear correlation of the variations in the two datasets. \textcolor{black}{We note, however, that the TIMED/SEE intensity drops to the lowest value just after the time of the TORUS-A exposure, where also the lowest value was found in the COS data.} We also note that the HST/COS brightnesses in the first two exposures, which are taken close in time at the off-target position 'E' and the torus-centered position 'C' (first two points in top panel of Figure \ref{fig:lya}), are almost identical and thus not indicative of contributions from a Europa torus.

We set an upper limit for the H Lyman-$\alpha$ torus brightness to 50~R. This limit corresponds to 10-times the difference in the first two exposures in 2020 and about $\sim$6 times the statistical uncertainty of the COS data. \textcolor{black}{This might appear to be a more conservative upper limit (compared to the OI limit), but ultimately the actual contributions of the geocorona to the signal in each exposure are not known. The assumed 50~R limit corresponds to the range in total brightness in the 2021 exposures (where the solar Lyman-$\alpha$ flux was more stable).}  

\begin{table}[htb]
\caption{HST/COS upper limits and model brightnesses for neutral emissions from electron impact and resonant scattering (atoms only).}
\label{tab:results2}
\begin{tabular}{lccccccccccc}
\hline
& \multicolumn{1}{c}{TORUS-B/C/D} &   & \multicolumn{5}{c}{Neutral tori model brightnesses (9.3~R$_\mathrm{J}$)} \\
Line    &   B$_{HST}$ &   &    & B$_{X,scat}$  $^{a}$  & B$_{X,el.imp.}$ $^{a}$ & B$_{X_2,el.imp.}$ $^{a}$ & B$_{tot}$\\
HI~1216~Å  &   $<$50~R   &    &  &   8~R  &  -- &  $<$1~R  &   8~R \\
OI~1304~Å  &   $<$0.6~R &    &  &   0.1~R  &  0.7~R  & $<$0.1~R &   0.8~R \\
OI~1356~Å  &   $<$0.6~R &    &  &   --    &  0.1~R  & $<$0.1~R  &   0.1~R \\

%\hline
%& \multicolumn{1}{c}{HST/COS - TORUS-A (2021)}    &   &  \multicolumn{5}{c}{Io torus model brightnesses (8.0~R$_\mathrm{J}$)} \\
%Line    &   B$_{HST}$ &   &    &  & B$_{X,el.imp.}$  \\
%SIII~1198~Å & 5.7$\pm$0.4~R \\
%SIV~1255~Å & 2.5$\pm$0.3~R  \\
%SII~1415~Å & 3.0$\pm$0.5~R  \\
\hline
& \multicolumn{1}{c}{TORUS-A}    &   &  \multicolumn{5}{c}{Neutral tori model brightnesses (8.0~R$_\mathrm{J}$)} \\
Line    &   B$_{HST}$ &   &    & B$_{X,scat}$ $^{a}$  & B$_{X,el.imp.}$  $^{a}$ & B$_{X_2,el.imp.}$  $^{a}$ & B$_{tot}$\\
%Line    &   B$_{HST}$ &   &    & B$_{X,scat}$  & B$_{X,el.imp.}$ & B$_{X_2,el.imp.}$ & B$_{tot}$\\
HI~1216~Å  &   $<$50~R   &    &  &   3-8~R  &  -- &  $<$1~R  &   4-9~R \\
OI~1304~Å  &   $<$0.6~R &    &  &   $<$0.1~R  &  0.7~R  & $<$0.1~R & 0.7~R\\
OI~1356~Å  &   $<$0.9~R &    &  &   --    &  0.1~R  &   $<$0.1~R &   0.1~R\\
\hline
\end{tabular}
\tablenotetext{a}{X refers to the respective atomic species, H or O, and X$_2$ to the molecular species, H$_2$ or O$_2$. "Scat" is the brightness only from resonant scattering, "el.imp" denotes the brightness from electron impact excitation.}
\end{table}

\section{Modeling and interpretation}

In order to relate the observational constraints to expected brightnesses and ion and neutral abundances, we use two models: First, we apply a 3-D Monte Carlo neutral torus model in order to reproduce predicted neutral particle column densities along observational lines of sight. For the sulfur ion detections, we use an Io plasma torus model with prescribed radial profiles for densities and temperatures of electrons and ions.   

\subsection{Neutral torus model}

 The neutral torus model is well validated and has been used extensively to study Saturnian and Jovian neutral tori \citep{Smith2005,Smith2007,Smith2008,Smith2010,smith19,Smith2021,Smith2022}. The 3-D Monte Carlo particle tracking computational code is capable of simulating hundreds of thousands of test particles over a period of years. During each model time step, groups of representative (weighted) particles are ejected from Io and Europa. These particle trajectories are tracked and influenced by gravitational fields of Jupiter and the Galilean satellites. The model accounts for particle interaction processes including electron impact ionization and dissociation, photo-ionization and photo-dissociation, recombination, charge exchange, neutral-neutral collisions, collision with Jupiter and the satellites as well as escape from the Jovian system. These particles then provide three-dimensional Jupiter neutral particle distributions for parent and daughter species (SO$_2$, SO, S, H$_2$O, OH, O, O$_2$, H and H$_2$). For particles that are dissociated, the original particle is removed and the resulting daughter particles are added.

\begin{figure}[htb]
\begin{center}
\includegraphics[width=0.95\textwidth]{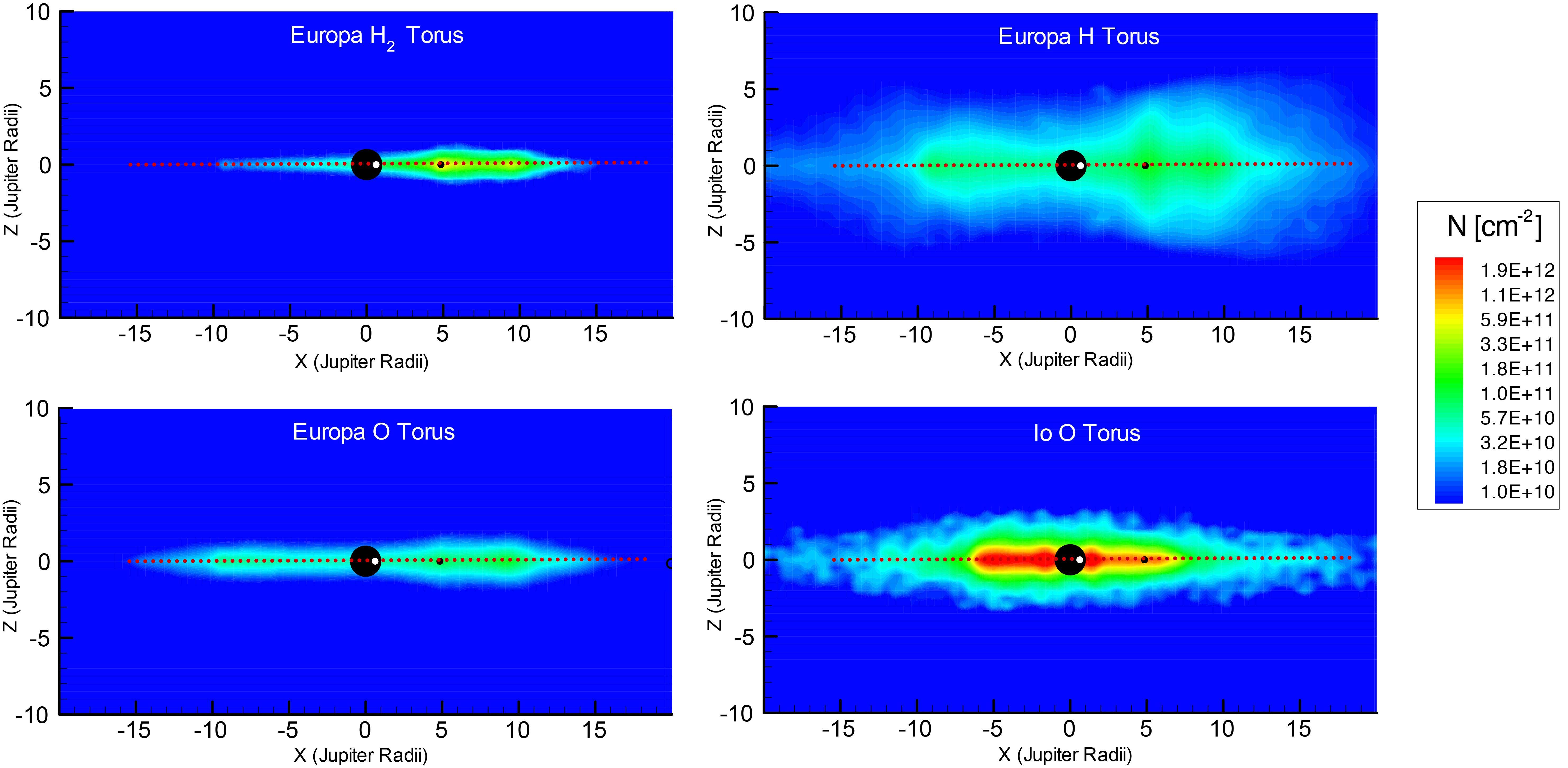}
\caption{Column density maps for H$_2$ (top left), H (top right), and O (bottom) from the torus model \citep{smith19,Smith2022} in a plane through Jupiter and corresponding to the viewing geometry of the first torus-centered exposure le2oa1010 (Table \ref{tab:obs}) from 2020. For O, the tori of Europa-genic (left) and Io-genic (right) material are shown separately. Jupiter (large black circle) is centered and Io (white, behind Jupiter) and Europa (black) are shown by the smaller disks to the right (west of Jupiter). The densities are higher closer to the moons (here, west of Jupiter) and the HST/COS observations scanned over the densest regions of the Europa-genic neutrals from 8.0~R$_\mathrm{J}$ to 12.5~R$_\mathrm{J}$. (Note that the viewing geometry in Figure \ref{fig:obsgeom} corresponds to an 1.5-hour earlier exposure).
\label{fig:smithmodel}}
\end{center}
\end{figure}

For this research, we updated the average particle interaction calculations based on \cite{smith19} and \cite{Smith2022}, which were derived from a survey of Galileo Plasma Science (PLS) plasma moments (in the NASA Planetary Data System) where the spacecraft was within 1$^\circ$ of Jupiter’s equatorial plane when no satellites were in the vicinity. These results provided a radial distribution of plasma (electrons and various ions) densities and temperatures for average conditions with densities and relative ion abundances determined by \cite{Steffl2004b}, \cite{smyth11}, \cite{bagenal15}, and \cite{yoshioka2017} and compared/adjusted for consistency with \cite{Bagenal2016}. The other key model inputs were Io and Europa particle source characterizations. For Europa, we utilized the source velocities and rates as defined by \cite{smyth06}, which includes source rates for H$_2$ (17 kg/s), O$_2$ (16 kg/s) and O (7 kg/s). For Io, we applied the source velocities from \cite{smyth03} and source rates for SO$_2$ (400 kg/s) and O (200 kg/s) based on \cite{Smith2022} and \cite{koga19}. We executed our model until convergence, which produced a dynamic 3-D magnetospheric distribution of neutral particle species. This model was then aligned with the correct geometry for each exposure to reproduce the observational lines-of-sight. Figure \ref{fig:smithmodel} shows the line-of-sight integrated column densities for different species from the Europa source and for O with Io as the source. \textcolor{black}{For reference, at 9.3~R$_\mathrm{J}$ (our TORUS-C position) the total (from both Io and Europa) line-of-sight column densities for H, H$_2$, O, and O$_2$ in this configuration are $N_\mathrm{H} = 2 \times 10^{11}$~cm$^{-2}$, $N_\mathrm{H_2} = 6 \times 10^{11}$~cm$^{-2}$, $N_\mathrm{O} = 3 \times 10^{11}$~cm$^{-2}$, and $N_\mathrm{O_2} = 2 \times 10^{10}$~cm$^{-2}$.}

With the simulated 3-D distributions, we estimate the emissions from resonant scattering of sunlight by H and O, as well as from electron impact (dissociative) excitation for O, O$_2$, H and H$_2$.  For resonant scattering we calculate scattering g values \citep[e.g.,][]{killen17,roth2023} for the two visits based on the solar flux and the distance to the Sun, see Table \ref{tab:solar_g}.  Column densities for H and O at the aperture positions are taken from the line-of-sight maps for H and O for each exposure (example in Figure \ref{fig:smithmodel}).
\begin{table}[htb]
\caption{Solar flux at 1~AU, heliocentric distance of Jupiter, and g values for the H and O FUV lines.}
\label{tab:solar_g}
\begin{tabular}{lcccccc}
\hline
Date     &   Solar Ly-$\alpha$ & Sun distance & g (HI 1216 Å)  & g (OI 1304 Å) \\
(UTC)   &   [cm$^{-2}$s$^{-1}$] &  [AU] & [s$^{-1}$]  & [s$^{-1}$] \\ 
2020-08-29 & $3.6\times10^{11}$   &   5.13  &  $6.0\times10^{-5}$  & $4.9\times10^{-7}$\\
2021-08-29 & $4.1\times10^{11}$   &   5.01  &  $7.4\times10^{-5}$  & $6.2\times10^{-7}$\\
\hline
\end{tabular}
\end{table}

Electron impact excitation is calculated using the same electron density and temperature profiles used for the neutral torus simulations. Electron impact excitation rates based on cross sections from \cite{ajello91,kanik03} and \cite{johnson03} are used. We first calculate the volume emission rate in each 3D grid point and integrate this along the line-of-side to obtain a surface brightness comparable to the observed normalized brightnesses.    

Calculations are carried out for each of the 10 exposures separately, considering the exact geometry (moon orbital positions) and aperture pointing. For the model values in the top part of Table \ref{tab:results2} for comparison to the TORUS-B,C,D exposures, the brightness is estimated for a column located at 9.3~R$_\mathrm{J}$ west of Jupiter (corresponding to TORUS-C position; Figure \ref{fig:obsgeom}). The modelled brightnesses for positions TORUS-B and TORUS-D are about 20\% to 40\% lower, respectively. For the TORUS-A exposures, the model brightness is calculated for the radial distance of 8.0~R$_\mathrm{J}$. Because the COS spectra are corrected with TORUS-E spectra, we model the brightnesses for this position (12.5~R$_\mathrm{J}$) and subtract these from the on-target brightness, to be consistent with the data correction. 

The model brightnessses most relevant for the data comparison are shown in Table \ref{tab:results2}. Taken together the model brightnesses for H Lyman-alpha are at least a factor of 6 below the sensitivity of the observations. The main contribution to H Lyman-alpha is resonant scattering by atomic hydrogen. Electron impact dissociative excitation of H$_2$ contributes at a level of $\leq10$\%.

The modelled OI-1304~Å brightnesses is about 30\% higher than our upper limit of 0.6~R. The expected OI~1356~Å emissions, which relate to a semi-forbidden transition and are excited only by electron impact, are well below the observational constraint.  Both oxygen emissions (at 1304~Å and 1356~Å) originate primarily from atomic oxygen, excited both through electron impact and resonant scattering (at 1304~Å). Dissociative excitation of molecular oxygen, O$_2$, contributes only about 1\% according to the simulations.

\subsection{Io sulfur ion torus model}

For Io torus sulfur ion emissions, we estimate the emissions based on previous work and models \citep{bagenal15,yoshioka2017,Nerney2017}. %The exciting electrons are assumed to be Maxwellian distributed with a constant core temperature of 7~eV.
The intensities of the emissions from the detected sulfur ions, which are excited by electron collisions, are calculated using the CHIANTI database version 8.0 \citep{Dere1997,delZanna2015}. To obtain brightnesses comparable to the observations, we need to integrate along the line-of-sight through the plasma torus, with the tangential point being the radial distance according to each pointing (see Figure \ref{fig:obsgeom}). For the radial profile for electron density we use the profile obtained by \cite{Steffl2004b}, also given in equation 6 of \cite{Nerney2017}. Thus, we assume a radially symmetric distribution as we did for the modelling of the neutral torus. Finally, in order to match the observed integrated brightnesses at the emission multiplets, we adjust the fractional abundance of the different sulfur species relative to the electron abundance.   
The best matching sulfur ion fractions are summarized and compared to previous results in Table \ref{tab:S_ions}. The simulated emission line spectra are then corrected for the Doppler shift due to the rotational velocity of the plasma torus at the radial distance of 8.0~R$_\mathrm{J}$. Assuming full co-rotation, the plasma moves away at 100.6~km/s, resulting in a red shift of 0.4~Å at the Lyman-$\alpha$ wavelength. The shifted model spectra, adjusted to the spectral resolution of COS data of 0.1~Å, are shown in red in Figure \ref{fig:Sion_detect}.    

\begin{table}[htb]
\caption{Sulfur ion fractional number density relative to the electron density, in comparison to previous results near 8~R$_\mathrm{J}$ radial distance. The numbers shown here are taken from studies that used CHIANTI version 8 \citep{yoshioka2017,Nerney2017}.}
\label{tab:S_ions}
\begin{tabular}{lcccccc}
\hline
        & This study & Yoshioka 2017 &  Nerney 2017  & Nerney 2017\\
        & HST/COS    &   Hisaki              & Cassini UVIS & Voyager UVS\\
Ions    & 8.0~R$_\mathrm{J}$ & 8.0~R$_\mathrm{J}$ & 8.0~R$_\mathrm{J}$ & 8.25~R$_\mathrm{J}$ \\
\hline
S$^+$    & 0.036 $\pm$0.004    &   0.025 $\pm$0.004  &   0.032 $\pm$0.002 &  0.019 $\pm$0.02  \\
S$^{++}$ & 0.23 $\pm$0.02     &   0.18  $\pm$0.03   &   0.18 $\pm$0.01   &  0.12 $\pm$0.03   \\
S$^{+++}$& 0.075 $\pm$0.01     &   0.073 $\pm$0.012 &   0.049 $\pm$0.003 &  0.10 $\pm$0.02   \\
%S$^+$    & 0.032    &   0.025   &   0.032   &  0.019  \\
%S$^{++}$ & 0.13     &   0.18    &   0.18    &  0.12   \\
%S$^{+++}$& 0.061    &   0.073   &   0.049   &  0.10   \\
\hline
\end{tabular}
\end{table}

Finally, we calculated the expected sulfur ion brightnesses at 9.3~R$_\mathrm{J}$ to compare them to the non-detections in the 2021 TORUS-C exposures. The modelled brightnesses are 0.1~R for both the SII emissions at 1255~Å and the SIV emissions at 1417~Å, and 0.2~R for the SIII emissions (1198~Å). These brightnesses are well below (SII and SIV) and just below (SIII) the sensitivity of the COS data and our non-detections (see, e.g., bottom panel of Figure \ref{fig:O_nondetect}) consistent with the model torus and the detections at 8.0~R$_\mathrm{J}$.

\section{Discussion and Summary}

The HST/COS spectra provided upper limits for H Lyman-$\alpha$ and OI far-UV emissions in the Jovian equatorial plane at radial distances between 8.0~R$_\mathrm{J}$ and 10.0~R$_\mathrm{J}$ west of Jupiter. Our model results for the neutral emissions from both electron impact and resonant scattering by the atomic species predict lower emissions for a neutral torus as derived in \cite{smith19}, with one exception.

The modelled OI~1304~Å brightness from electron impact of O is about 30\% higher than the HST/COS upper limit on the emissions. This could indicate a lower abundance of O in the torus as assumed in \cite{smith19}. The abundance of O in the model is similar to abundances of O$_2$ and H and 2--4 times lower than the abundance of the primary Europa neutral torus species, H$_2$, around 9.3~R$_\mathrm{J}$ (cf. Figure \ref{fig:smithmodel}). Thus, a 30\% lower O abundance would hardly change the overall torus density, which means the \cite{smith19} model scenario would still be consistent with the ENA measurements \citep{mauk03,Mauk2004}. A lower abundance of O might be related to a lower source of escaping O from Europa, consistent with recent lower upper limits on the O abundance in Europa's bound atmosphere \citep{Roth2021-Eur}.

Even at the most inward COS position (8.0~R$_\mathrm{J}$), neutral abundances are predicted to be similar to the Europa torus center (near 9.3~R$_\mathrm{J}$). The model shows that most of the oxygen at this distance originates from Europa, and only inwards  of 7.0~R$_\mathrm{J}$ is neutral oxygen from Io more abundant than the Europa-genic oxygen (see bottom panels in Figure \ref{fig:smithmodel}). 

The recent results from the Juno mission \citep{Szalay2022} suggest an H$_2$ source rate of 1.2~kg/s for the torus, which is about five times lower than assumed in earlier studies \citep{shematovich05,smyth06} and more than an order of magnitude below the source rate assumed in our modelling. Another possibility is that the Europa neutral torus is overall more dilute than assumed in our model, which would lead to overall lower expected emissions, even less detectable by HST/COS. On the other hand, a recent study of Ganymede's H corona \citep{roth2023} suggests that also Europa's H corona is denser than derived in the study of the first detection \citep{roth17-europa}. How a denser atomic hydrogen corona can be consistent with the relatively low molecular hydrogen density in the torus \citep{Szalay2022} could be a subject of study for atmosphere simulations like the recent study by \cite{Carberry2023} for Callisto. 

We note, however, that electron impact excitation depends highly on the electron density and the electron temperature dependent cross sections, which have not been updated in over 20 years. Thus, the fact that the predicted OI~1304~Å brightness from electron impact on O is higher than the observational upper limit might simply be explained by the uncertainty in the electron excitation. Furthermore, the modelled brightnesses from electron impact on molecular oxygen O$_2$ are well below the observational limit, emphasizing that an O$_2$ cloud in Europa's orbit is very difficult to detect at these wavelengths. 
   
Our limit on the HI~1216~Å Lyman-$\alpha$ brightness of 50~R is about 6 times higher than expected from our modeling (Table \ref{tab:results2}), which might even overestimate hydrogen abundances. \textcolor{black}{For a signal of 50~R, a line of sight H column density of $N_\mathrm{H} >7 \times 10^{11}$~cm$^{-2}$ would be required well above the simulated abundances.} This shows how difficult it is to obtain a direct detection of the Europa neutral hydrogen torus at far-UV wavelengths. Ultraviolet measurements from spacecraft like the current Juno or upcoming JUpiter ICy Moon Explorer (JUICE) and Europa Clipper missions could search for neutral H but would require sufficiently long integration times.   

Finally, emissions from sulfur ions (S+, S++, and S+++) were detected in 2021 at a radial distance of 8.0~R$_\mathrm{J}$. Io torus emissions, including these sulfur ion emissions, were previously detected at similar radial distances to Jupiter by Voyager UVS \citep{Shemansky1988}, by Cassini UVIS during the Jupiter flyby \citep{Steffl2004b} and by the Hisaki telescope \citep{yoshioka2017}. Modeling the detected emission brightnesses with an Io torus model, we find abundances of the sulfur ions relative to the assumed electron abundance that agree well with previous measurements when comparing to the studies that used the same version (8.0) of the CHIANTI database for the analysis (Table \ref{tab:S_ions}). Small differences between the earlier measurements and our results can easily be explained by the measurement uncertainties, \textcolor{black}{time-variability in the torus \citep[e.g.,][]{delamere03}} or even differences in the pointing and aperture size. The non-detections of sulfur ion emissions at 9.3~R$_\mathrm{J}$ in the conjugate exposures is consistent with modelled lower brightnesses at the larger radial distance due to the lower plasma densities.   

The sulfur ion emissions were not detected at the same radial distance (8.0~R$_\mathrm{J}$) in 2020. A possible explanation for the non-detection can be the geometry. In 2020, the torus was seen at a slightly larger inclination towards the observer (Figure \ref{fig:obsgeom}) compared to 2021. This leads to a lower line-of-sight column density and thus lower brightnesses. A small offset of the small round aperture from the torus center might also lead to a drop in abundances and brightnesses below the detection threshold. Another possibility is of course the time-variability in torus, which might play some role, although there is no independent evidence for a dilute torus phase in 2020 to our knowledge. 

Taken together, the HST/COS far-UV scan of the equatorial region between 8.0~R$_\mathrm{J}$ and 10.0~R$_\mathrm{J}$ provided the lowest direct upper limits on hydrogen and oxygen abundances to date, which are, however, also consistent with most recent models and estimations for the Europa neutral torus as well as with the expected Io neutral torus. In addition, the observations provided one detection of sulfur ion emissions that confirms the sulfur ion fractionation in the extended Io torus derived from several earlier measurements.

%% Also note that the akcnowlodgment environment does not support long amounts of text. If you have a lot of people and institutions to acknowledge, do not use this command. Instead, create a new \section{Acknowledgments}.

\begin{acknowledgments}
L.R. appreciates support from the Swedish National Space Agency through grant 2021-00153. This research is based on observations made with the NASA/ESA Hubble Space Telescope obtained from the Space Telescope Science Institute, which is operated by the Association of Universities for Research in Astronomy, Inc., under NASA contract NAS 5–26555. These observations are associated with program 15848, with appreciated support to N.C., T.B., K.R., and M.V. 
All of the data presented in this paper were obtained from the Mikulski Archive for Space Telescopes (MAST) at the Space Telescope Science Institute. The specific observations analyzed can be accessed via\dataset[10.17909/9074-ez94]{http://dx.doi.org/10.17909/9074-ez94}.
\end{acknowledgments}


\vspace{5mm}
\facilities{HST(COS)}




%\appendix

%\section{Appendix information}


\bibliography{bib2023}{}
\bibliographystyle{aasjournal}



\end{document}